\documentclass[a4paper,11pt]{article}
\pdfoutput=1 

\usepackage{jheppub} 

\usepackage[T1]{fontenc} 

\usepackage{amsfonts,amsmath,amssymb,mathrsfs}
\usepackage{enumerate}
\usepackage{hyperref}
\usepackage{graphicx}
\usepackage{caption}
\usepackage{subcaption}
\usepackage{ytableau}

\usepackage{enumitem}  

\usepackage{comment} 
    
\newcommand{\be}{\begin{equation}}
\newcommand{\ee}{\end{equation}}

\newcommand{\bea}{\begin{eqnarray}}
\newcommand{\eea}{\end{eqnarray}}

\newcommand{\nn}{\nonumber}

\newcommand{\dd}{\textrm{d}}
\newcommand{\AdS}{\textrm{AdS}}

\newcommand{\cc}[1]{{\bf \textcolor{purple} {{[CC: #1]}}}}

\title{\boldmath  Gluing two discs into a spindle}

\author[a]{Hyojoong Kim}
\author[b]{and Nakwoo Kim}

\affiliation[a]{Department of Physics and Astronomy \& Center for Theoretical Physics, \\
Seoul National University, Seoul 08826, Korea}
\affiliation[b]{Department of Physics 
and Research Institute of Basic Science,  Kyung Hee University,\\ 
Seoul 02447, Korea}

\emailAdd{hjkim1996@gmail.com}
\emailAdd{nkim@khu.ac.kr}

\abstract{We construct AdS$_5 \times \Sigma$ solutions of $D=7$ gauged supergravity where $\Sigma$ is a disc, also known as a half-spindle, and uplift them to $D=11$ supergravity. By computing the anomaly polynomial of the M5-brane theory compactified on $\Sigma$ in the large $N$ limit, we obtain the off-shell central charges of the dual $D=4$, $\mathcal{N}=1$ superconformal field theory to disc solutions. These off-shell central charges decompose into contributions from the center and the boundary of the disc and are referred to as the gravitational blocks for disc. By suitably gluing the gravitational blocks for two discs along their boundaries, we reproduce the off-shell central charge for a single spindle. }

\begin{document} 
\maketitle
\flushbottom

\noindent
\section{Introduction}
Supersymmetric solutions of supergravity theories that contain 
an AdS factor have provided a natural arena for exploring holography via the AdS/CFT correspondence. Over the past few years, a new class of AdS$_d \times \Sigma$ solutions of various gauged supergravity theories have been constructed for $d=2,\ldots, 5$, where the two dimensional geometries $\Sigma$ are known as {\it spindles} \cite{Ferrero:2020laf} and {\it discs} \cite{Bah:2021mzw,Bah:2021hei}. A spindle is topologically a two sphere with conical singularities at the north and south poles. A disc can be regarded as a ``half-spindle'' with a conical singularity at the center and a curvature singularity at the boundary of the disc. (See \cite{Ferrero:2020twa, Hosseini:2021fge, Boido:2021szx, Ferrero:2021wvk, Couzens:2021rlk,Faedo:2021nub, Ferrero:2021etw, Giri:2021xta,Couzens:2021cpk} for spindles and \cite{Suh:2021ifj, Suh:2021aik, Suh:2021hef, Couzens:2022yjl,Bah:2022yjf} for discs.)

These gravity solutions suggest that the dual superconformal field theories in $d-1$ dimensions can be obtained by compactifying the $d+1$ dimensional field theories dual to the parent AdS$_{d+2}$ solution on a spindle or a disc, and flowing to the IR.
The identification of superconformal field theories dual to spindle geometries is currently a subject of intensive studies \cite{Inglese:2023wky,Inglese:2023tyc,Colombo:2024mts}, see also \cite{Pittelli:2024ugf, Mauch:2024uyt, Jeon:2025vtu}. The field theories dual to $\mathcal{N}=2$ disc solutions in $D=7$ gauged supergravity have class $\mathcal{S}$ interpretations as M5-branes wrapped on a sphere with a regular and an irregular puncture \cite{Bah:2021mzw,Bah:2021hei}. 

Physical observables such as the central charge and the free energy can be easily calculated for given gravity backgrounds after uplifting the lower dimensional solutions to $D=10$ or $11$ supergravity. Recently, it has been shown that their off-shell expressions can be decomposed into basic building blocks, referred to as {\it gravitational blocks}. 
The notion of gravitational blocks was first proposed in \cite{Hosseini:2019iad} for AdS black holes and black strings with spherical horizons, and was applied to spindles in \cite{Hosseini:2021fge, Faedo:2021nub}. Especially, in \cite{Faedo:2021nub}, the authors conjectured that the off-shell free energies/central charges of SCFTs which can be obtained from the branes wrapped on a spindle can be written in a gravitational block form for M2, D3, D4 and M5-branes. By gluing two gravitational blocks, i.e. the contributions from the north and the south pole of a spindle, and extremizing it, one can obtain the on-shell free energy/central charges. The gravitational block formulas can be derived for M2 and D3-branes within the framework of GK geometries \cite{Boido:2022mbe}.
Finally, thanks to the recent studies on equivariant localization in the context of AdS/CFT correspondence, the gravitational blocks can be understood as
contributions from the fixed point set by using the Berline–Vergne–Atiyah–Bott (BVAB) fixed point formula \cite{BenettiGenolini:2023kxp, Martelli:2023oqk, BenettiGenolini:2023yfe,BenettiGenolini:2023ndb, Colombo:2023fhu}.  

In this paper, we will study $\mathcal{N}=1$ (1/4-BPS) disc solutions of $D=7$ gauged supergravity theory and their uplift to $D=11$ supergravity. 
The disc solutions can be understood as a ``half-spindle'' and since the spindle solutions preserve $\mathcal{N}=1$ supersymmetry, it is more appropriate to study $\mathcal{N}=1$ disc solutions rather than the well-known $\mathcal{N}=2$ disc solutions. Computing the anomaly polynomials for M5-branes wrapped on a disc in the large $N$, we will show that the off-shell central charges of $D=4$, $\mathcal{N}=1$ SCFT dual to disc solutions can be decomposed into two parts: the contributions from the center and the boundary of a disc. The latter can be understood as a new class of the gravitational blocks\footnote{As discussed in the main text, we suspect that there may be additional contributions from the boundary.} and studied from the perspective of the equivariant localization with boundaries \cite{Couzens:2024vbn,Cassani:2024kjn,Colombo:2025ihp}. 

Finally, we will consider summing up the off-shell central charges for two discs, where the orientation of the second disc is opposite to that of the first. We will show that the contributions from the boundaries of the first and second discs exactly cancel, once we require that the periods of the two discs are equal. Then, the remaining contributions from the centers of two discs give rise to the off-shell central charge for a single spindle with a certain identification of parameters. We interpret this process as gluing two discs into a single spindle.

This paper is organized as follows. In section \ref{N=1 disc}, we study AdS$_5 \times \Sigma$ solutions of $D=7$ gauged supergravity, focusing on $\mathcal{N}=1$ disc or half-spindle solutions. In section \ref{uplift}, we uplift the solutions to $D=11$ supergravity and calculate the holographic central charge $a$. Section \ref{GB} is devoted to the investigations on the gravitational block expression for $\mathcal{N}=1$ disc. Then, we consider gluing two discs and derive the off-shell central charge $a$ for a single spindle from glued discs in section \ref{gluing}.
We conclude with a discussion in section \ref{discussion}.
\section{M5-branes on an $\mathcal{N}=1$ disc}\label{N=1 disc}
In this section, we study supersymmetric AdS$_5 \times \Sigma$ solutions of seven-dimensional maximal $SO(5)$ gauged supergravity truncated to the $U(1) \times U(1)$ sector.
The local solutions we are interested in are given by  \cite{Ferrero:2021wvk}
\begin{align}
\dd s_7^2&= \left( y\, P(y) \right)^{1/5}
\left(\dd s_{\AdS_5}^2+\dfrac{y}{4\,Q(y)}\dd y^2+\dfrac{Q(y)}{P(y)}\dd  z^2 \right),\\
A_i &=\left(-1+\dfrac{q_i}{h_i(y)}\right) \dd z, \quad X_i(y)=\dfrac{(y P(y))^{2/5}}{h_i(y)},
\end{align}
where
\begin{align}
&h_i(y)=y^2+q_i, \\
&P(y)= \left( y^2+q_1 \right)\left( y^2+q_2 \right),\\
&Q(y)= -y^3+\dfrac{1}{4}
\left( y^2+q_1 \right)\left( y^2+q_2 \right).
\end{align}
Here $q_i$'s are two real parameters and we use the gauge adopted in \cite{Ferrero:2021etw}.
From the same local solution depending on the roots of the quartic equation $Q(y)=0$, we obtain various solutions which we quickly summarize. Let us denote the real roots by $y_\alpha$. 
If there are four real roots, one can obtain the well-known $\mathcal{N}=1$ spindle solution by taking $0<y_2<y<y_3$ \cite{Ferrero:2021wvk}. If there is a double root at $y_1=0$, the solutions with $0<y<y_2$ reduce to the $\mathcal{N}=2$ disc \cite{Bah:2021mzw,Bah:2021hei,Couzens:2022yjl}. When there are two real roots ($0<y_1<y_2$), solutions with $y_2<y<\infty$ correspond to domain wall solutions \cite{Gutperle:2022pgw}. 

One class of solution that has so far not been studied, are the solutions with $0<y<y_1$. As we will see these seem to be the $\mathcal{N}=1$ cousins of the $\mathcal{N}=2$ disc solutions. 
At $y=y_1>0$, the polynomial $Q(y)$ vanishes. Then, the circle parameterized by $z$ shrinks and there can be a conical singularity. At the other end-point at $y=0$ we have $P(y=0)=4Q(y=0) \neq 0$. Then, there is a circle with a finite radius and the space looks like a cylinder. Note that the warp factor of AdS$_5$ shrinks at $y=0$. Then the space is conformally $\text{AdS}_5\times \mathbb{R}\times S^1$ at this end-point. Hence this solution corresponds to a disc having an orbifold singularity at the center and a curvature singularity at the boundary.\footnote{$\mathcal{N}=1$ disc solutions were studied in $D=6,\, F(4)$ gauged supergravity in \cite{Suh:2021aik} and in $D=7$ gauged supergravity coupled with three vector multiplets in \cite{Karndumri:2022wpu}. See also appendix C of \cite{Couzens:2022lvg}.} 

Now let us focus on the explicit two-dimensional metric for a disc,
\be
\dd s_\Sigma^2= \dfrac{y}{4Q(y)}\dd y^2 +\dfrac{Q(y)}{P(y)}\dd z^2.
\ee
As $y$ approaches $y_1$, the metric takes the form
\be
\dd s_\Sigma^2 \simeq\dfrac{y_1}{|Q'(y_1)|}\left(\dd r^2+\dfrac{|Q'(y_1)|^2}{4y_1^4} r^2 \dd z^2\right),
\ee
where $r^2=y_1-y$. We demand the period of $z$ to be
\be
\dfrac{Q'(y_1)}{2y_1^2} \Delta z=-\frac{2\pi}{\ell},
\ee
with positive integer $\ell$. Then, the metric has an orbifold singularity at $y=y_1$ with the conical deficit angle $2\pi (1-\frac{1}{\ell})$. 

The Euler number may be computed as
\begin{align}
\chi(\Sigma)&=\dfrac{1}{4\pi}\int_{\Sigma} R_\Sigma \textrm{vol}_\Sigma
=\dfrac{1}{4\pi}\int_{\Sigma} 2 \left(\dfrac{Q P'-Q' P}{y^{1/2}P^{3/2}}\right)' \dd y \dd z\nn\\
&=-\dfrac{\Delta z}{2\pi} \dfrac{Q'(y_1)}{2 y_1^2}=\dfrac{1}{\ell}.
\end{align}
In general, there is an additional contribution from the boundary however this vanishes since the intrinsic curvature of the boundary vanishes. This also occurs when considering the $\mathcal{N}=2$ disc.

The magnetic charges are given by
\be
P_i \equiv \dfrac{1}{2\pi} \int_\Sigma \dd A_i =-\dfrac{\Delta z}{2\pi} \dfrac{y_1^2}{y_1^2+q_i}.
\ee
Then we obtain
\be\label{mag-ch}
P_1+P_2= \dfrac{1}{\ell}-\dfrac{3}{2}\dfrac{\Delta z}{2\pi}.
\ee
We demand that the magnetic charges are quantized as
\be
P_i=\dfrac{p_i}{\ell}, \quad\quad p_i \in \mathbb{Z}.
\ee
The R-symmetry gauge field $A^R \equiv A_1 +A_2$ satisfies
\be
A^R\Big|_{y=y_1}= \left(-\dfrac{3}{2}\dfrac{\Delta z}{2\pi} +\dfrac{1}{\ell}\right) \dd \phi,
\quad\quad A^R\Big|_{y=0}=0,
\ee
at the center and the boundary of the disc where we define a new angular coordinate $\phi \equiv \frac{2\pi}{\Delta z} z$. 
Let us consider the gauge transformation
\be\label{GT}
\tilde{A}^{(I)} \equiv A^{(I)}+s^{(I)} \dd z,
\ee
with the constraint $s^{(1)} +s^{(2)} =\frac{3}{2}$, then the spinors are independent of the $z$ coordinate \cite{Ferrero:2021etw}. In this gauge, we have
\be
\tilde{A}^R\Big|_{y=y_1}= \dfrac{1}{\ell} \dd \phi,
\quad\quad \tilde{A}^R \Big|_{y=0}=\dfrac{3}{2}\dfrac{\Delta z}{2\pi} \dd \phi.
\ee

Observe that the solutions are uniquely characterized by three integers: $p_1,p_2$ and $\ell$. We can express the roots  $y_1$, the parameters $q_i$ and the period $\Delta z$ of the disc solutions in terms of this data as
\begin{equation}
\begin{split}\label{N=1 disk sol}
q_1 &=-\dfrac{27\, p_1\, p_2^2\,(p_1-2\,p_2+2)}{(1-p_1-p_2)^4}\, ,\quad q_2 =-\dfrac{27\, p_1^2\, p_2\,(p_2-2\,p_1+2)}{(1-p_1-p_2)^4}\, ,\\
y_1 &=\dfrac{9\, p_1\, p_2}{(1-p_1-p_2)^2}\, ,\quad\qquad\qquad \Delta z =\dfrac{2(1-p_1-p_2)}{3 \ell}2\pi\, .
\end{split}
\end{equation}
Note that we need to impose two conditions on the $p$'s. Firstly we have $y_1>0$ which implies $p_1 p_2>0$ and thus both are of the same sign. Moreover, we require $\Delta z>0$ and therefore $1\geq p_1+p_2$. This fixes both magnetic charges to be negative.

\section{11d uplift and the central charge}\label{uplift}


We now want to study the solution directly in $D=11$ supergravity. The $D=7$ gauged supergravity can be consistently uplifted to $D=11$ supergravity on an $S^4$. Using the uplift we will study the regularity of the metric before quantizing the fluxes and computing the central charge.

\subsection{Uplift to M-theory}
We uplift the seven-dimensional solution to $D=11$ supergravity by following \cite{Cvetic:1999xp}. Then the eleven-dimensional metric is written as
\be
L^{-2}\dd s_{11}^2=\Omega^{1/3} \dd s_7^2 +\Omega^{-2/3}\left[X_0^{-1} \dd \mu_0^2
+\sum_{i=1}^2 X_i^{-1} \left(\dd \mu_i^2+\mu_i^2\left(\dd \phi_i+A_i \right)^2\right)\right],
\ee
with 
\be
X_0 \equiv (X_1 X_2)^{-2}, \quad \Omega \equiv \sum_{a=0}^2 X_a \mu_a^2, \quad \sum_{a=0}^2 \mu_a^2=1.
\ee
Here the coordinates $\mu_a$ and $\phi_i$ parametrize the internal $S^4$.
We also parametrize the $\mu_a$ as
\be
\mu_0= \mu, \quad \mu_1= \sqrt{1-\mu^2}\cos\theta,\quad \mu_2= \sqrt{1-\mu^2}\sin\theta.
\ee
The curvature singularity at $y=0$ of a disc solution of $D=7$ gauged supergravity acquires a physical interpretation in the uplifted solutions. Near $y=0$, the solution is given by
\begin{align}
\dd s_{11}^2& \approx \dfrac{1}{4} \left(q_1\,q_2 \right)^{1/3} \mu^{2/3} y^{-1/3} \left(4 \dd s_{\AdS_5}^2+\left(q_1 \, q_2\right)^2 \dd z^2 \right)\\
&+y^{2/3} \left(\dfrac{\mu}{q_1 q_2}\right)^{2/3} \left( \dd y^2
+  \dfrac{q_1 \cos^2\theta+q_2 \sin^2\theta}{1-\mu^2}\dd\mu^2
+\dfrac{q_1-q_2}{\mu}\sin2\theta \dd\mu \dd\theta \right.\nn\\
&\phantom{+y^{2/3} \left(\dfrac{\mu}{q_1 q_2}\right)^{2/3}\left(\right.}\left. +\dfrac{1-\mu^2}{\mu^{2}}
 \left(q_1 \cos^2\theta D\phi_1^2+q_2 \sin^2\theta D\phi_2^2 
 +\left(q_1 \sin^2\theta+q_2\cos^2\theta \right)\dd\theta^2\right)\right),\nn
\end{align}
where $D\phi_i=\dd \phi_i+A_i$.
This describes M5-branes wrapped on AdS$_5\times S^1$. 
To identify this singularity it is useful to recall that the metric for a stack of M5-brane solutions smeared along $s$-dimensional subspace of the transverse space can be written as
\begin{align}
\dd s_{11}^2&= H^{-1/3}\left(-\dd t^2+\dd\vec{x}_5^2 \right) 
+H^{2/3}\left(\dd r^2+r^2 d\Omega_{4-s}^2+\dd\vec{y}_s^2 \right)\, ,\\
&H=1+\dfrac{Q}{r^{3-s}}\, .
\end{align}
Note that our solution can be reproduced by setting $s=4$ and ignoring the integration constant $1$ in the harmonic function. 
Thus, it can be interpreted as M5-brane solutions smeared along the $\mu, \theta, \phi_1$ and $\phi_2$  directions.
This type of brane singularity has appeared previously in \cite{Bah:2015fwa} where it was called an M9-brane and is assumed to be the prototype of the uplift of a D8-brane to 11d supergravity.\footnote{We thank Chris Couzens for bringing the reference \cite{Bah:2015fwa} to our attention.}

\subsection{Central charge}
Now let us calculate the central charge of the $D=4$, $\mathcal{N}=1$ SCFT dual to M5-branes wrapped on a disc. Given a $D=11$ metric of the form
\be
\dd s_{11}^2=L^2 e^{2\lambda} \left(\dd s_{\AdS_5}^2+\dd s_{M_6}^2 \right),
\ee
the central charge of the dual SCFT \cite{Gauntlett:2006ai} is 
\be
a=\dfrac{1}{2^7 \pi^6} \left(\dfrac{L}{\ell_p}\right)^9 \int_{M_6} \dd ^6 x \sqrt{g_{M_6}} e^{9\lambda}.
\ee
Once we identify the warp factor and the internal metric as
\begin{align}
e^{2\lambda}&= \Omega^{1/3} \left(y P(y) \right)^{1/5},\\
\dd s_{M_6}^2 &= \dfrac{y}{4Q(y)}\dd y^2 +\dfrac{Q(y)}{P(y)}\dd z^2+
\Omega^{-1} \left(y P(y) \right)^{-1/5}
\left[
X_0^{-1} \dd \mu_0^2
+\sum_{i=1}^2 X_i^{-1} \left(\dd \mu_i^2+\mu_i^2\left(\dd \phi_i+A_i \right)^2\right)\right],\nn
\end{align}
we obtain
\be
\sqrt{g_{M_6}} e^{9\lambda}=\dfrac{1}{4}\, y\,(1-\mu^2)\sin2\theta.
\ee
Then, the central charge $a$ can be easily calculated to be
\be
a=\dfrac{1}{2^7 \pi^6} \left(\dfrac{L}{\ell_p}\right)^9 \int_{\Sigma} \dfrac{y}{2}\,\dd y\, \dd z \, \textrm{vol}(S^4)
=\dfrac{1}{2^7 \pi^6} \left(\dfrac{L}{\ell_p}\right)^9 \dfrac{8\pi^2}{3} \dfrac{y_1^2}{4} \Delta z ,
\ee
where vol$(S^4)=\frac{8\pi^2}{3}$ and the ranges of the angles are $\mu \in [-1,1],\,\theta \in [0,\frac{\pi}{2}]$ and $\phi_i \in[0,2\pi]$.
From the flux quantization of the four-form flux, we have
\be
L= (\pi N)^{1/3} \ell_p,
\ee
and finally we obtain the central charge
\be\label{a-grav}
a=\dfrac{9\, p_1^2\, p_2^2}{16\, \ell\, (1-p_1-p_2)^3}N^3\, .
\ee
Observe that the central charge is positive only for $(1-p_1-p_2)>0$, which is precisely the same condition we found from our $D=7$ analysis for the period to be positive.

\section{Gravitational blocks for discs}\label{GB}
The gravitational blocks, originally introduced for AdS black holes in \cite{Hosseini:2019iad} and studied for spindles in \cite{Hosseini:2021fge, Faedo:2021nub}, are the basic building blocks to calculate the physical quantities such as central charges and entropy. One can glue the gravitational blocks evaluated at the north and the south poles of a spindle and extremize it to obtain the on-shell values of the physical observables. In this section, we study the gravitational block expression of an off-shell central charge for M5-branes wrapped on a disc.
\subsection{Field theory calculation : anomaly polynomial}
Our strategy for obtaining the gravitational block expression for a disc is to use the M5-brane anomaly polynomial and reduce this on the disc. For M5-branes wrapped on a spindle, the anomaly polynomial was used to obtain the central charge $a$ in \cite{Ferrero:2021wvk,Martelli:2023oqk}. Here, we will briefly summarize the procedure for the readers' convenience (see also \cite{Hosseini:2020vgl}). In the large $N$ limit, the M5-brane anomaly polynomial is given by
\be
\mathcal{A}_{6\textrm{d}}=\dfrac{1}{24}p_2(R) N^3,
\ee
where $p_2(R)$ is the second Pontryagin class of the $SO(5)_R$ normal bundle. 
For the Cartan subgroup $U(1) \times U(1) \subset SO(5)_R$, the eight-form reduces to
\be\label{A6d}
\mathcal{A}_{6\textrm{d}}=\dfrac{1}{24}c_1(\mathcal{N}_1)^2c_1(\mathcal{N}_2)^2 N^3,
\ee
where $c_1(\mathcal{N}_i)$ are the first Chern classes of the complex line bundle $\mathcal{N}_i$. Now let us consider compactification of the worldvolume theory on $\Sigma$. Then the eight-dimensional auxiliary space $Z_8$ where $\mathcal{A}_{6\textrm{d}}$ is defined is the total space of a $\Sigma$-fibration over the six-dimensional auxiliary space $Z_6$ as
\be
\Sigma \hookrightarrow Z_8 \rightarrow Z_6.
\ee
The first Chern class $c_1(\mathcal{N}_i)$ is decomposed as
\be\label{1stCC}
c_1(\mathcal{N}_i)= \Delta_i c_1(R_{4\dd})-c_1(\mathcal{L}_i),
\ee
where $\Delta_i$ is the trial R-charge satisfying $\Delta_1+\Delta_2=2$ and $R_{4\dd}$ is the pullback of a $U(1)_R$ symmetry bundle over $Z_6$. The second term is the first Chern class 
$c_1(\mathcal{L}_i)=[\mathscr{F}_i/2\pi]\in H^2(Z_8,\mathbb{R})$ where we introduce connection one-forms on $Z_8$ as\footnote{Here we used the gauge considered in \eqref{GT} with the specific value $s^{(1)}=s^{(2)}=3/4$. }
\be
\mathscr{A}_i=\left(\dfrac{q_i}{h_i(y)}-\dfrac{1}{4}\right) \dfrac{\Delta z}{2\pi} \left(\dd \phi +\mathcal{A_J}\right)
\equiv \rho_i(y) (\dd \phi+\mathcal{A_J}).
\ee 
Note that here $\mathcal{A_J}$ is introduced due to the $\Sigma$-fibration over $Z_6$.
Then, the curvature becomes
\be
\mathscr{F}_i =\rho'_i(y) \dd y\wedge (\dd \phi+\mathcal{A_J})+\rho_i(y) \mathcal{F_J},
\ee
where $\mathcal{F_J}\equiv \dd \mathcal{A_J}$ with the first Chern class $c_1(\mathcal{J})=[\mathcal{F_J}/2\pi]\in H^2(Z_6,\mathbb{R})$.
Now inserting the first Chern class \eqref{1stCC} into the $d=6$ anomaly polynomial \eqref{A6d} and integrating it over $\Sigma$ as
\be\label{a-4d}
\mathcal{A}_{4\textrm{d}}=\int_\Sigma \mathcal{A}_{6\textrm{d}},
\ee
one can get the $d=4$ anomaly polynomial. As a final step, by formally setting $c_1(\mathcal{J})=\epsilon\, c_1(R_{4\dd})$ and reading off the coefficient of $c_1(R_{4\dd})^3$ in $\mathcal{A}_{4\textrm{d}}$, we obtain the central charge $a$
\begin{align}\label{a-spindle}
a_{\textrm{trial}}=\dfrac{9\times 6 }{32}\dfrac{N^3}{24}
\Big[&-\Delta_1\Delta_2(\Delta_1 I_1+\Delta_2 I_2)
+(\Delta_1^2 I_3+ 2 \Delta_1 \Delta_2 I_4+ \Delta_2^2 I_5)\,\epsilon\nn\\
&-(\Delta_1 I_6+\Delta_2 I_7)\,\epsilon^2+I_8\, \epsilon^3\Big].
\end{align}
The explicit expressions of $I_i$, which are certain integrals of $\rho_i(y)$ and $\rho'_i(y)$, can be found in the appendix of \cite{Ferrero:2021wvk}.

Now let us compute the central charge for a disc, using the prescription summarized above. The formal expression of the trial central charge for a disc is exactly the same as that of a spindle \eqref{a-spindle}, except for the explicit form of $\rho_i(y)$ and the integration ranges.
Here, for a disc, we have $\rho_i(y)$ at $y=y_1$ and $y=0$ to be
\begin{align}
&\rho_1(y_1)=\dfrac{1+p_1-p_2}{2\ell},\quad  \rho_1(0)=\dfrac{1-p_1-p_2}{2\ell},\nn\\
&\rho_2(y_1)=\dfrac{1-p_1+p_2}{2\ell} , \quad   \rho_2(0)=\dfrac{1-p_1-p_2}{2\ell}.
\end{align} 
We managed to write down the trial central charge for a disc as
\be\label{a_disc_trial}
a_{\textrm{trial}}^{\textrm{disc}}=\dfrac{9 }{128} N^3\dfrac{1}{\epsilon}\left( \left(\Delta_1-\epsilon \rho_1 \right)^2
 \left(\Delta_2-\epsilon \rho_2  \right)^2-\left(\Delta_1 \Delta_2 \right)^2\right)\bigg|^{y_1}_{0}.
\ee
Note that, in contrast to the spindles, the trial central charge for a disc is not maximized to yield the correct central charge, which agrees with the gravitational calculation \eqref{a-grav}. Instead, if we use the relation between the extremal value of $\epsilon$ and the period $\Delta z$ reported in \cite{Ferrero:2021wvk},
\be
\epsilon^*=\dfrac{4}{3}\dfrac{2\pi}{\Delta z},
\ee
and insert this value into \eqref{a_disc_trial} with $\Delta z$ computed in \eqref{N=1 disk sol}, then we observe that the correct central charge is obtained. (Here, we set $\Delta_1=\Delta_2=1$ for simplicity.)

\subsection{Off-shell central charge of disc}
Now let us rewrite the central charge for a disc \eqref{a_disc_trial} in a more concise form as
\be\label{cc-gb-2}
a_{\textrm{trial}}^{\textrm{disc}}=\dfrac{9}{16}
\dfrac{\left(b_1^{\textrm{c}}\, b_2^{\textrm{c}}\right)^2-\left(b_1^\textrm{b}\, b_2^\textrm{b}\right)^2}{\epsilon/2} N^3,
\ee
where
\be
b_i^{\textrm{c}}\equiv \dfrac{1}{2}\Big(\Delta_i -\rho_i(y_1)\epsilon \Big),\quad
b_i^{\textrm{b}}\equiv \dfrac{1}{2}\Big(\Delta_i -\rho_i(0)\epsilon \Big).
\ee
We use the superscripts c and b to denote the contributions to the central charge $a_{\textrm{trial}}$ evaluated at the center ($y=y_1$) and the boundary ($y=0$) of the disc, respectively. More explicitly, the form of $b_i^{\textrm{c}}$ and $b_i^{\textrm{b}}$ are 
\bea
b_1^{\textrm{c}}&=&\dfrac{1}{2}\left(\Delta_1-\dfrac{1+p_1-p_2}{2\ell}\epsilon \right), \quad
b_2^{\textrm{c}}=\dfrac{1}{2}\left(\Delta_2-\dfrac{1-p_1+p_2}{2\ell}\epsilon \right),\nn\\
b_1^{\textrm{b}}&=&\dfrac{1}{2}\left(\Delta_1-\dfrac{1-p_1-p_2}{2\ell}\epsilon \right),\quad
b_2^{\textrm{b}}=\dfrac{1}{2}\left(\Delta_2-\dfrac{1-p_1-p_2}{2\ell}\epsilon \right).
\eea
As a result, the central charge for a disc can be decomposed into two parts, i.e. the contributions from the center and the boundary of a disc, respectively. It is precisely the gravitational block expression for a disc we are looking for.
Note that even though the form of the trial central charge for a disc \eqref{cc-gb-2} is the same to that of the spindle \eqref{a-spindle-gb}\footnote{One can also check that $b_i^{\textrm{c}},\,b_i^{\textrm{b}}$ satisfy the following relations
\be
b_1^{\textrm{c}}+b_2^{\textrm{c}}=1-\dfrac{1}{\ell}\dfrac{\epsilon}{2},\quad
b_1^{\textrm{b}}+b_2^{\textrm{b}}=1-\dfrac{1-p_1-p_2}{\ell}\dfrac{\epsilon}{2},
\ee
where $1-p_1-p_2>0$. Compared to the spindle case \eqref{b-spindle},
this implies that the topology is totally different from the spindle.}, it is not extremized over $\epsilon$ and $\Delta_i$ as we already mentioned in the previous section. Instead, if we require that there are no boundary contributions to the on-shell central charge, we need to impose $b_i^{\textrm{b}}=0$. Then we find $\Delta_i^*=1,\, \epsilon^*=2\ell/(1-p_1-p_2)$ and successfully reproduce the central charge \eqref{a-grav} obtained in the gravity calculation. From now on, we call the trial central charge \eqref{cc-gb-2} the off-shell central charge for a disc in gravitational block form.

One might suspect that there are additional boundary terms, which do not contribute to the on-shell value of the central charge, but are necessary for $a$-maximization to work.
For example, let us consider the following term, without any physical interpretation,
\be
a_{\textrm{bdy}}= \dfrac{1}{48} \left(
-\Delta_1 \Delta_2\left( \Delta_2\,q_1
+\Delta_1\,q_2 \right)\rho(y)
+\left(\Delta_2^2\,q_1 
+\Delta_1^2\,q_2\right)\rho(y)^2\epsilon
\right)\bigg|_{y=0}N^3,
\ee
and add it to the central charge $a_{\textrm{trial}}^{\textrm{disc}}$ \eqref{cc-gb-2}.
This term leads the total central charge to be extremized with the constraint $\Delta_1+\Delta_2=2$ to give the extremal values $\Delta_i^*=1,\, \epsilon^*=2\ell/(1-p_1-p_2)$. As a result, the gravity result can be successfully reproduced by $a$-maximization. At this moment, the derivation of the non-trivial additional boundary contributions $a_{\textrm{bdy}}$ from the $d=6$ anomaly polynomial is not clear. These terms might originate from computing the $d=4$ anomaly polynomial in \eqref{a-4d} due to the presence of the non-trivial boundary of the disc. We leave this issue for future work.   

\section{Gluing two discs}\label{gluing}
Now let us consider gluing two discs together to form a spindle from the perspective of the gravitational blocks. First, we add the off-shell central charges \eqref{cc-gb-2} of the two discs as\footnote{In this section, we omit the superscript `disc' for brevity.}
\bea\label{cc-fin-1}
\mathfrak{a} &\equiv&
a_{\textrm{trial}}+\hat{a}_{\textrm{trial}},\nn\\
&=&\dfrac{9}{16}
N^3
\left(\frac{\Big(b_1^{\textrm{c}}\, b_2^{\textrm{c}}\Big)^2-\Big(b_1^{\textrm{b}}\, b_2^{\textrm{b}}\Big)^2}{\epsilon/2}
+\frac{\Big(\hat{b}_1^{\textrm{c}}\, \hat{b}_2^{\textrm{c}}\Big)^2-\Big(\hat{b}_1^{\textrm{b}}\, \hat{b}_2^{\textrm{b}}\Big)^2}{\hat{\epsilon}/2}\right).
\eea
The hatted quantities belong to the second disc with the opposite orientation, which implies that we have $\hat{\epsilon}=-\epsilon$ and $\hat{p}_1+\hat{p}_2>1$. Here we assumed $\Delta_i=\hat{\Delta}_i$. Second, we require that the periods of two discs we are going to glue are identical, i.e. $\Delta z=\Delta \hat{z}$. In terms of the disc data, this requirement is equivalent to
\be\label{cons}
\dfrac{1-p_1-p_2}{\ell}=-\dfrac{1-\hat{p}_1-\hat{p}_2}{\hat{\ell}}.
\ee
Under this condition, one can show that $b_i^\textrm{b}=\hat{b}_i^\textrm{b}$ and, in turn, the off-shell boundary contributions of two discs in \eqref{cc-fin-1} are exactly canceled.\footnote{Recently, a similar idea was used in \cite{Colombo:2025ihp} to illustrate the boundary contributions in equivariant localization in odd dimensions. The authors have provided a simple example for computing the volume of the round three sphere. First, the volume of a hemisphere can be decomposed into two parts i.e. the contributions from the center and from the boundary of hemisphere, respectively. Second, by summing up the volumes of northern and southern hemisphere, the boundary contributions cancel and the volume of $S^3$ is obtained by summing up the contributions from the centers of hemispheres.  
} As a result, the total off-shell central charge of the glued discs becomes
\bea\label{cc-tot-discs}
\mathfrak{a}&=&\dfrac{9}{16}
\dfrac{N^3}{\epsilon/2}
\left(\Big(b_1^{\textrm{c}}\, b_2^{\textrm{c}}\Big)^2
-\Big(\hat{b}_1^{\textrm{c}}\, \hat{b}_2^{\textrm{c}}\Big)^2\right),\nn\\
&=&
 \dfrac{9 }{16} \dfrac{N^3}{\epsilon/2}
\left[ \left(\dfrac{1}{2}\left(\Delta_1-\dfrac{1+p_1-p_2}{2\ell}\epsilon \right)\dfrac{1}{2}\left(\Delta_2-\dfrac{1-p_1+p_2}{2\ell}\epsilon  \right)\right)^2\right.\nn\\
&\phantom{=}&\phantom{ \dfrac{9 }{16} \dfrac{N^3}{\epsilon/2}}- \left.  \left(\dfrac{1}{2}\left(\Delta_1+\dfrac{1+\hat{p}_1-\hat{p}_2}{2\hat{\ell}}\epsilon \right)\dfrac{1}{2}\left(\Delta_2+\dfrac{1-\hat{p}_1+\hat{p}_2}{2\hat{\ell}}\epsilon  \right)\right)^2\right].
\eea
We note that $b_i^{\textrm{c}}$ and $\hat{b}_i^{\textrm{c}}$ satisfy the following relations
\bea\label{b-disc}
b_1^\textrm{c} +b_2^\textrm{c}=1-\dfrac{1}{\ell}\dfrac{\epsilon}{2},\quad\quad
b_1^\textrm{c}-\hat{b}_1^\textrm{c}=-\dfrac{\hat{\ell}\,p_1+\ell\, \hat{p}_1}{\ell \,\hat{\ell}}\dfrac{\epsilon}{2},
\nn\\
\hat{b}_1^\textrm{c} +\hat{b}_2^\textrm{c}=1+\dfrac{1}{\hat{\ell}}\dfrac{\epsilon}{2},\quad\quad
b_2^\textrm{c}-\hat{b}_2^\textrm{c}=-\dfrac{\hat{\ell}\,p_2+\ell\, \hat{p}_2}{\ell \,\hat{\ell}}\dfrac{\epsilon}{2},
\eea
which enables us to identify the glued discs as a spindle. 
For a spindle, the trial central charge can be written as
\be\label{a-spindle-gb}
\mathfrak{a}^{\textrm{spindle}}=\dfrac{9[  \left(\mathfrak{b}_1^+ \mathfrak{b}_2^+\right)^2- \left(\mathfrak{b}_1^- \mathfrak{b}_2^-\right)^2]}{ 16\varepsilon}N^3.
\ee
Here, $\mathfrak{b}_i^\pm$ called the weights satisfy the following relations
\be\label{b-spindle}
{\mathfrak{b}_1^\pm}+{\mathfrak{b}_2^\pm} =1\mp\frac{\varepsilon}{n_{\pm}}, \quad 
\mathfrak{b}_i^+ -\mathfrak{b}_i^-=-\frac{\mathfrak{p}_i}{n_+ n_-}\varepsilon,
\ee
where the fibration structure is encoded \cite{BenettiGenolini:2023kxp,BenettiGenolini:2023ndb}.
Comparing \eqref{b-disc} and \eqref{b-spindle}, we identify our variable $b_i^\textrm{c}$ and $\hat{b}_i^\textrm{c}$ with the weights
\be
\mathfrak{b}_i^+ \equiv b_i^\textrm{c}, \quad \mathfrak{b}_i^- \equiv \hat{b}_i^\textrm{c}.
\ee
We further identify the parameters and charges of the spindle with those of two discs
\bea\label{map}
&&n_+ \equiv \ell, \quad \quad n_- \equiv \hat{\ell}\nn\\
&&\mathfrak{p}_i \equiv \ell\, \hat{p}_i +\hat{\ell}\,p_i,
\quad \varepsilon \equiv\epsilon/2.
\eea
One can check that the constraint for the spindle, $\mathfrak{p}_1+\mathfrak{p}_2=n_++n_-$, is satisfied once we use \eqref{cons}.
Now we follow the same procedures described in \cite{BenettiGenolini:2023ndb} and introduce the new variables $\Phi_i$
\be
\mathfrak{b}_i^\pm =\dfrac{1}{2} \left(\Phi_i \mp \dfrac{\mathfrak{p}_i}{n_+ n_-}\varepsilon\right),
\ee 
with the constraint
\be
\Phi_1+\Phi_2 = 2+\dfrac{n_+-n_-}{n_+ n_-}\varepsilon.
\ee
As a result, we reproduce the off-shell central charge for a single spindle \cite{Faedo:2021nub,Martelli:2023oqk,BenettiGenolini:2023ndb}
\bea\label{cc-tot-spindle}
\mathfrak{a}
&=&-\dfrac{9
\left(\Phi_1 \mathfrak{p}_2+\Phi_2 \mathfrak{p}_1\right)
\left( \mathfrak{p}_1\mathfrak{p}_2 \varepsilon^2+
n_+^2 n_-^2 \Phi_1 \Phi_2\right)}
{64\,n_+^3\,n_-^3}N^3,
\eea
from the off-shell central charge for the glued discs \eqref{cc-tot-discs}. 

Let us summarize the results we discussed in this section. From the expression \eqref{cc-fin-1}, we can calculate the central charges for two disjoint discs or a single spindle from the glued discs, under certain conditions, respectively. If we require that there are no boundary contributions to the on-shell central charge for a disc, i.e. if we impose $b_i^{\textrm{b}}=\hat{b}_i^{\textrm{b}}=0$, then total central charge becomes
\be
\mathfrak{a}=\dfrac{9\, p_1^2\, p_2^2}{16\, \ell\, (1-p_1-p_2)^3}N^3+\dfrac{9\, \hat{p}_1^2\, \hat{p}_2^2}{16\, \hat{\ell}\, (1-\hat{p}_1-\hat{p}_2)^3}N^3.
\ee
It is just a sum of central charges for the first disc and the second with the opposite orientation.
On the other hand, if we require that the two discs have the same period, we can impose $b_i^{\textrm{b}}=\hat{b}_i^{\textrm{b}}$. Then the total central charge for the glued discs \eqref{cc-tot-discs} reduces to that of a single spindle \eqref{cc-tot-spindle}. In this case, the on-shell central charge is determined by $a$-maximization.

Spindles have four parameters $(n_+,n_-,\mathfrak{p}_1,\mathfrak{p}_2)$ with one constraint $\mathfrak{p}_1+\mathfrak{p}_2=n_++n_-$. On the other hand, glued discs have six parameters $(\ell, \hat{\ell},p_1,p_2,\hat{p}_1,\hat{p}_2)$ with one constraint $(1-p_1-p_2)/\ell=-(1-\hat{p}_1-\hat{p}_2)/\hat{\ell}$. 
We make a few comments on the charges of the spindle and the glued discs in \eqref{map}.
First, the magnetic charges of the spindle are invariant under shifts of the disc magnetic charges as $p_i \rightarrow p_i-k_i\, \ell,\,\hat{p}_i \rightarrow \hat{p}_i+k_i\, \hat{\ell}$, with $k_i \in \mathbb{Z}$.\footnote{A similar analysis is carried out in the gluing of two local models near the orbifold points to obtain a spindle in \cite{Ferrero:2021etw}.}
Second, by using the gluing constraint \eqref{cons}, the magnetic charges of the spindle can be rewritten as
\be
\mathfrak{p}_1= \ell+\hat{\ell}-\left(\ell\, \hat{p}_2 +\hat{\ell}\,p_2 \right),\quad
\mathfrak{p}_2=\ell\, \hat{p}_2 +\hat{\ell}\,p_2.
\ee
This implies that, for the glued discs, we have 
\be
\mathfrak{p}_1 \mathfrak{p}_2<0,\quad \textrm{or}\quad \mathfrak{p}_1>0,\,\mathfrak{p}_2>0.
\ee
However, an explicit supergravity spindle solution is known only for the case $\mathfrak{p}_1 \mathfrak{p}_2<0$ in \cite{Ferrero:2021wvk}.  

Let us consider spindles and glued discs with the specific values of the parameters. For example, the central charge for the spindle with $(n_+=1,\,n_-=2,\,\mathfrak{p}_1=-1,\,\mathfrak{p}_2=4)$ is maximized to give $\mathfrak{a}=2/75N^3$ with $\Phi_1^*=1/5,\, \varepsilon^*=-2/15$. Using the dictionary between the parameters of spindle and disc \eqref{map}, this spindle corresponds to the glued discs with $(\ell=1,\, \hat{\ell}=2,\,p_1\,,p_2,\,\hat{p}_1=-1-2p_1,\,\hat{p}_2=4-2p_2)$. Then, by $a$-maximization, we obtain $\mathfrak{a}=2/75N^3$ at $\Delta_1^*=-2/15(p_1-p_2), \, \varepsilon^*=-2/15$ while $p_i$'s are undetermined. Since we have $0 \leq\Delta_1^* \leq 2$, we have to impose further restrictions on $p_i$, i.e. $-15 \leq p_1-p_2 \leq 0$ in this example. 
This shows that a two-parameter family of glued discs yields the same central charge as the corresponding spindle with specific parameters.

\section{Discussion}\label{discussion}
In this paper we have studied $\mathcal{N}=1$ disc solutions of $D=7$ gauged supergravity, which have a conical singularity at the center and a curvature singularity at the boundary of a disc. By computing M5-brane anomaly polynomials in the large $N$, we have decomposed the off-shell central charge $a$ of $D=4$, $\mathcal{N}=1$ SCFT dual to disc solutions into two parts i.e. the contributions from the center and the boundary of a disc, respectively.
The former plays a similar role as the gravitational blocks studied in AdS black holes and spindles. On the other hand, the latter is an entirely new class of gravitational blocks, which originates from the boundary of a disc.
With this gravitational block expression of central charge for a disc, we have successfully derived the off-shell central charge for a single spindle by gluing two discs along the boundaries. Here gluing refers to adding the off-shell central charges for the first disc and the second disc with  opposite orientation, under the constraint that the periods of the two discs are equal.

For $\mathcal{N}=2$ disc solutions, it is well established that they are holographic duals of M5-branes wrapped on a sphere with one regular and one irregular puncture, which correspond to an orbifold singularity at the center and a curvature singularity at the boundary of disc, respectively \cite{Bah:2021mzw,Bah:2021hei,Couzens:2022yjl,Bah:2022yjf,Bomans:2023ouw,Couzens:2023kyf}. Spindle solutions were also considered from class $\mathcal{S}$ viewpoint in \cite{Bomans:2024mrf}. In that perspective, gluing two $\mathcal{N}=1$ discs along the boundaries we have studied in this paper corresponds to the gluing of two irregular punctures in dual field theory. Hence it would be interesting to understand class $\mathcal{S}$ interpretations of this gluing process. Furthermore, gluing two spindles at their poles, i.e. gluing the south pole of the first spindle and the north pole of the second and identifying its class $\mathcal{S}$ interpretation as gluing of two regular punctures are also very interesting topics for future studies.%

  As we already pointed out in the main text, the on-shell value of the central charge for a single disc cannot be obtained by $a$-maximization. Computing anomaly polynomials and central charges associated to the manifolds with boundaries might give non-trivial contributions. It would be interesting to figure out these additional boundary terms, which vanish on-shell and make $a$-maximization work. Equivariant localization with boundary \cite{Couzens:2024vbn,Cassani:2024kjn,Colombo:2025ihp} may offer a resolution to this problem. It could also be worthwhile to investigate the properties of boundaries, incorporating field-theoretic studies of partition functions on manifolds with boundaries, as for example studied for three-dimensional SCFTs in \cite{Beem:2012mb,Yoshida:2014ssa,Dimofte:2017tpi}. We hope to address these issues in the near future.


\section*{Acknowledgments}
We are grateful to Chris Couzens for the collaboration at the early stage of the project and the helpful discussions and comments throughout this work.
We would like to thank Heng-Yu Chen, Hee-Joong Chung, Seok Kim, Aaron Poole, Augniva Ray, Jaewon Song, Sang-Heon Yi for useful discussions. This research was supported by Basic Science Research Program through the National Research Foundation of Korea (NRF) funded by the Ministry of Education RS-2023-00243491 (HK) and RS-2025-00518906 (NK). HK acknowledges the hospitality of the string theory group at National Taiwan University, where part of this work was completed. 

\bibliography{disc}{}
\end{document}